\documentclass{aa}
\usepackage{graphicx}

\newcommand{\HI}{{\ion{H}{1}}}

\newcommand{\kms}{$\,$km$\,$s$^{-1}$}

\newcommand{\microJy}{$\mu$Jy beam$^{-1}$ }
\newcommand{\msun}{{$M_\odot$}}

\def\HI{H{\,\small I}}

\def\emph#1{{\sl #1}}
\newcommand{\ltsima} {$\; \buildrel < \over \sim \;$}
\newcommand{\gtsima} {$\; \buildrel > \over \sim \;$}
\newcommand{\lta} {\lower.5ex\hbox{\ltsima}}
\newcommand{\gta} {\lower.5ex\hbox{\gtsima}}

\input{psfig}

\begin{document}

\title{A Deep WSRT 1.4 GHz Radio Survey of the Spitzer Space Telescope
  FLSv Region}

\titlerunning{}
\authorrunning{Morganti et al.}

\author{R.  Morganti\inst{1}, M.A. Garrett\inst{2}, S. Chapman \inst{3}
W. Baan\inst{1}, G. Helou\inst{3}, T. Soifer\inst{3}}

\offprints{morganti@astron.nl}

\institute{Netherlands Foundation for Research in Astronomy, Postbus 2,
7990 AA, Dwingeloo, The Netherlands
\and
Joint Institute for VLBI in Europe, Postbus 2,
7990 AA, Dwingeloo, The Netherlands 
\and
California Institute of Technology, MS 320-47, Pasadena, CA 91125
}

\date{Received ...; accepted ...}

\abstract{ 
The First Look Survey (FLS) is the first scientific product to emerge from the
Spitzer Space Telescope.  A small region of this field (the verification
strip) has been imaged very deeply, permitting the detection of cosmologically
distant sources.  We present Westerbork Synthesis Radio Telescope (WSRT)
observations of this region, encompassing a $\sim 1$ sq. deg field, centred on
the verification strip (J2000 RA=17:17:00.00, DEC=59:45:00.000). The radio
images reach a noise level of $\sim 8.5$ \microJy\ - the deepest WSRT image
made to date.  We summarise here the first results from the project, and
present the final mosaic image, together with a list of detected sources. 
The
effect of source confusion on the position, size and flux density of the
faintest sources in the source catalogue are also addressed.  The results of a
serendipitous search for \HI\ emission in the field are also presented.  Using
a subset of the data, we clearly detect \HI\ emission associated with four
galaxies in the central region of the FLSv. These are identified with nearby,
massive galaxies.
}
\maketitle
\section{Introduction}

The Spitzer Space Telescope's (formerly SIRTF, the Space Infrared Telescope
Facility) First-Look Survey (FLS, http://ssc.spitzer.caltech.edu/fls/), is
expected to go two orders of magnitude deeper than any previous, large-area
Mid-IR survey.  The extragalactic component of the survey is centred on a
region that is within the continuous viewing zone of the satellite, in an area
of low IR cirrus foreground emission. The survey is composed of two parts - a
large-area shallow survey covering an area of $2.5^{\circ} \times 2^{\circ}$
and a smaller $\sim 0.75^{\circ} \times 0.3^{\circ}$ strip, referred to as the
``verification'' survey (or FLSv). The FLSv observations lie within the
shallow survey region, and are centred at RA=17:17:00.00, DEC=59:45:00.000,
J2000. By employing integration times $\sim 10$ times longer than the shallow
survey, the FLSv will reach largely unexplored (mJy and sub-mJy) sensitivity
levels in the mid-IR.  It is expected that cosmologically distant star forming
galaxies will form the dominant population of the faint FLSv sources counts.

The results of the FLS are expected to rapidly enter the public domain, and
they are supported by various space and ground-based ``follow-up''
observations. Parallel observations of the FLS in different wave-bands are
crucial in order to fully exploit the new and unique infrared data coming from
Spitzer. Complimentary optical, radio and sub-mm data have already been (or
are in the process of being) collected. Deep radio observations of the field
are of particular interest, since it is well known that for local star forming
galaxies, there exists a close correlation between their far-IR and radio
luminosities (e.g.  Helou et al. 1985, Condon 1992). This tight correlation
has recently been demonstrated to also apply at cosmological redshifts and
mid-IR wavelengths (e.g. Garrett 2002, Elbaz et al. 2002).  At sub-mJy levels,
various deep radio surveys (e.g. Richards et al.  2000) demonstrate that the
radio source counts are dominated by star forming galaxies located at moderate
redshifts ($z\sim 0.7$). There is also significant overlap between the
faintest $\mu$Jy radio sources and the cosmologically distant ($z \sim 2-4$)
sub-mm (rest frame FIR) population detected in deep, blank field SCUBA
observations (e.g.  Chapman et al. 2002).

A 1.4 GHz radio survey of the entire FLS region has been made by Condon
et al. (2003) using the VLA in the B-array configuration.  The final
survey image has a resolution of 5 arcseconds and reaches an average
($1\sigma$) rms noise level of 23\microJy.  A catalogue of 3565
radio components with peak flux densities greater than 115 $\mu$Jy ($5
\sigma$) has been compiled and is publicly available at
http://www.cv.nrao.edu/sirtf\_fls/. 

Fig. 1 shows the observed radio/sub-mm/FIR Spectral Energy Distribution
(SED) of the well known Ultra-luminous IR Galaxy, Arp 220, projected to
$z \sim 0.3, 0.7$, and 1.1. The 5-sigma sensitivity levels probed by
both the VLA and Spitzer FLS and FLSv observations are also presented.
The VLA 1.4 GHz survey (Condon et al.  2003) is able to detect Arp 220
type systems out to about $z\sim 0.5$.  The shallow Spitzer observations
go deeper, capable of detecting Arp 220 out to $z\sim 0.7$. The Spitzer
FLSv observations employ much longer integrations times, and are able
to go a factor of 3 times deeper (in terms of r.m.s. noise level) at 24
and 70 micron (160 micron observations are expected to be confusion
limited). The Condon et al.  (2003) VLA radio survey, is a relatively
poor match to the Spitzer observations, especially within the deeper area
of the FLS verification strip.

\begin{figure} 
\centerline{\psfig{figure=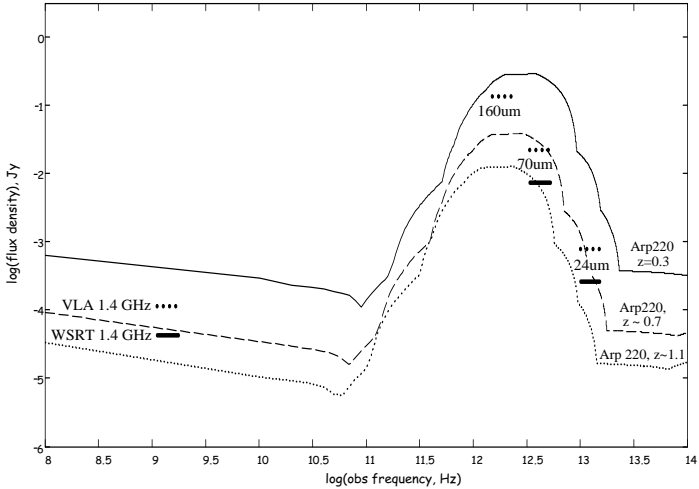,angle=-90,width=9.5cm}}

\caption{The SED of Arp 220 (at radio, sub-mm
  and FIR frequencies) projected to various redshifts (z=0.3, 0.7 \&
  1.1). The 5-sigma detection threshold for both the WSRT and VLA 1.4
  GHz observations are presented. For comparison, the expected 5-sigma
  detection threshold levels of Spitzer at 24, 70 and 160 micron for both
  the shallow survey (dotted line) and deeper verification survey
  regions (solid lines) are also shown. }

\end{figure}

With a clear need for a much deeper radio survey, over a large area of sky, we
started on a project that used the upgraded Westerbork Synthesis Radio
Telescope (WSRT) to image a $\sim 1$ sq. deg field, centred on the FLSv
strip. The resulting images reach a ($1\sigma$) noise level of $\sim 8.5$
\microJy - the deepest WSRT image made to date. The 5-sigma detection
threshold for the WSRT 1.4 GHz observations presented here are shown in Figure
1. The WSRT observations can detect Arp 220 like systems out to about $z\sim
0.7$. The FLSv Spitzer observations should be able to detect Arp 220 out to
about $z\sim 1$.  Only hyper-luminous, star forming systems can be detected
beyond $z\sim 1$ by both the Spitzer Space Telescope, VLA and WSRT FLSv survey
observations.

Both the VLA and WSRT-FLSv catalogues provide accurate
sub-arcsecond, astrometric source positions. These are likely to play a
crucial role in the reliable identification of Spitzer sources at other
wavelengths. In addition, a comparison of the FIR and radio flux ratios
can be used to estimate redshifts for the sources (via the FIR/radio
correlation {\it e.g.} Carilli \& Yun, 2000) or for those sources with
measured redshifts, to  determine (unobscured) star formation rates and
to distinguish between AGN and star formation processes as the main source
of the radio and FIR emission.

In this paper, we describe the WSRT 1.4 GHz observations of the FLSv strip,
the analysis of the data and the construction of the corresponding (on-line)
source catalogue. We also present the first preliminary results of a
serendipitous search for \HI\ emission in the field.

\begin{figure}
\centerline{\psfig{figure=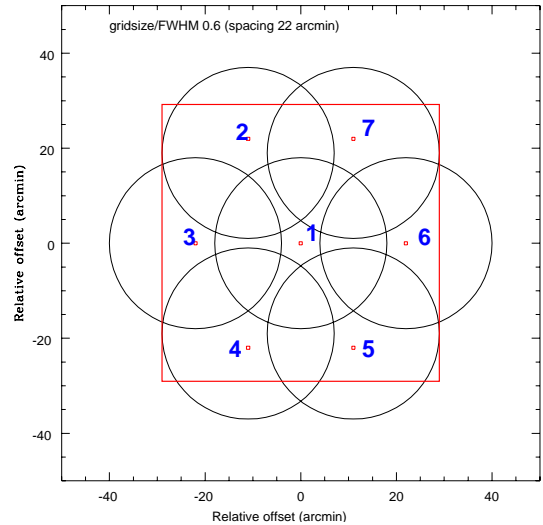,angle=0,width=7cm}}
\caption{Mosaic pattern of the 7 pointing centres used to image the FLSv. 
    The circles represent the FWHM of the WSRT primary beam. The
    central field is centred on RA(J2000)=17:17:00.00 and
    DEC(J2000)=59:45:00.000).  The square represents approximately the
    region used for the extraction of the source catalogue. The region
    goes from RA= 17:20:55 DEC=59:14:37 to RA= 17:13:00 DEC=60:14:37
    (see text for details). The rms noise rises from $\sim 8.5$ in the
    central area of the image, up to $\sim 20$ \microJy\ at the edge of the
    catalogued region.}
\end{figure}

\section{WSRT  observations and data reduction}

The total observing time originally requested for the WSRT observations of the
FLSv (156 hours) was based on the total areal coverage of the elongated FLSv
strip. The WSRT observations were made prior to the successful launch of the
Spitzer Space Telescope, and since the orientation of the FLSv strip was then
unknown, it was decided to image a circular region, centred on the the
FLSv. We chose a mosaic of 7 pointings, with a layout similar to that employed
by de Vries et al. (2002) in the WSRT observations of the NOAO-N (Bootes)
field. The position of the pointing centres for the WSRT FLSv observations are
shown in Fig. 2. The central field is centred on RA(J2000)=17:17:00.00 and
DEC(J2000)=59:45:00.000). The separation between the grid points was chosen to
be 60\% of the FWHM of the primary beam (the WSRT primary beam has a FWHM
$\sim 35$ arcminutes at 1.4 GHz).

\begin{figure*}
\vskip2cm
\centerline{Fig.3 (separate gif figure)}
\vskip2cm
\caption{{\sl Left} Image of the full WSRT mosaic of the FLSv. {\sl Right}
  The central 30$^\prime \times 30^\prime$ of the mosaic image.  The
  faintest sources visible in this image are at the level of 40 $\mu$Jy
  (peak flux density). }
\end{figure*}

The main observations were made during the period May-July 2002, about
16 hours was spent on each pointing. Two additional 12-hour
observations of the central region, were carried out in Feb 2003.

The upgraded WSRT is well suited to this type of deep, wide area
survey. The new back-end system is capable of imaging out the full
field of view of the WSRT antenna primary beam without distortion. The
very good sensitivity of the new Multi-Frequency Front End (MFFE) receiver
systems, ensures that a single 12 hour observation attains a noise level
approaching the expected confusion limit at 1.4 GHz.

We made use of the 160 MHz broad-band IF system available at the WSRT.  The
full band is made up of eight, independently tuned, 20~MHz bands.  The 8 bands
cover the frequency range between 1420 and 1280 MHz.  In the last two
observations of the central field, however, the 8 bands were centred between
1450 and 1311 MHz - this was found to be the optimum frequency range to use,
with respect to RFI considerations. For each of the 8 bands, 128 channels were
used (a total of 1024 channels were obtained for the full 160 MHz band) and
two polarisation products were recorded. Each 12 hr block was combined with
observations (before and after) of an intensity calibrator (typically
3C~147, 3C~48 and/or 3C~286).

The spectral-line mode in which the observations were carried out is
the standard mode of operation for continuum observations at the WSRT.
This ensures that bandwidth smearing effects are minimised when imaging
large fields of view, and also permits the rejection and isolation
of narrow RFI spikes, as well as foreground galactic emission.
Furthermore, it also permits an interesting by-product, i.e.  the
possibility of serendipitously detecting \HI\ emission from galaxies in
the field (see \S 5).

The data were calibrated and reduced using the MIRIAD software package
(Sault et al. 1995). Every pointing was mapped using a multi-frequency
synthesis approach, where the measurements of all the individual channels
are gridded simultaneously in the uv-plane. All the fields were imaged
using uniform weighting. Typically three steps of phase
self-calibration were employed, and the CLEAN image deconvolution was
performed down to the 3-$\sigma$ noise level. In the final
deconvolution, frequency dependent effects of
the primary beam shape were taken into account (Sault \& Wieringa
1994). All the images were restored with the same synthesised beam of
$14^{\prime\prime} \times 11^{\prime\prime}$ (p.a. = 0$^\circ$).  The
final maps were corrected for the primary beam response and linearly
combined in a mosaic. 

\begin{table*}
\centering
\caption{Example of the radio catalogue. The complete 
table of 714 sources can be found in electronic format in the WSRT-FLSv page
available at http://www.astron.nl/wsrt/WSRTsurveys/WFLS.  The individual
columns are as follows: Right ascension (hours, minutes, seconds), Declination
(degree, arcminutes and arcseconds), integrated flux
density (mJy), major and minor axes (in arcsecond) and position angle (in
degrees), local background rms (mJy, see text for details).  Major and minor
axes and position angle of the sources are not deconvolved for the synthesised
beam.}
\begin{tabular}{lcccccc}
\hline\hline\\
       RA   &
       DEC  & 
 Flux density &
$\Theta_{maj}$ &
$\Theta_{min}$ & 
PA  & rms$_{backg}$    \\
(J2000)  & 
(J2000) & mJy & arcsec & arcsec & degrees & mJy  \\
\hline \\
17:20:55.927 & 59:26:15.26 &   0.267 & 12.8 & 11.2 & -4.8 & 0.0155 \\
17:20:55.311 & 59:26:45.79 &   0.260 & 17.9 & 12.9 &-74.8 & 0.0195 \\
17:20:55.850 & 59:35:41.65 &   0.171 & 16.7 & 12.3 &-40.8 & 0.0189 \\
17:20:57.189 & 59:54:45.97 &   0.101 & 15.4 & 10.3 & 25.3 & 0.0149 \\
17:20:56.790 & 59:56:46.29 &   0.116 & 19.6 & 12.3 & -9.6 & 0.0132 \\
17:20:50.551 & 59:15:12.81 &   1.630 & 14.7 & 11.2 & 10.5 & 0.0273 \\
17:20:55.200 & 59:49:50.67 &   0.131 & 16.4 & 11.0 &-24.1 & 0.0157 \\
17:20:53.402 & 59:41:16.88 &   0.130 & 29.8 & 11.5 &  2.5 & 0.0132 \\
17:20:53.421 & 59:44:18.04 &   0.091 & 15.5 & 11.4 &-24.9 & 0.0149 \\
17:20:50.753 & 59:32:55.72 &   0.233 & 17.4 & 13.5 & 30.9 & 0.0163 \\
17:20:52.516 & 59:48:10.41 &   0.224 & 15.7 & 13.1 & -0.4 & 0.0139 \\
\hline \\
\end{tabular}
\end{table*}

\begin{figure*} 
\centerline{\psfig{figure=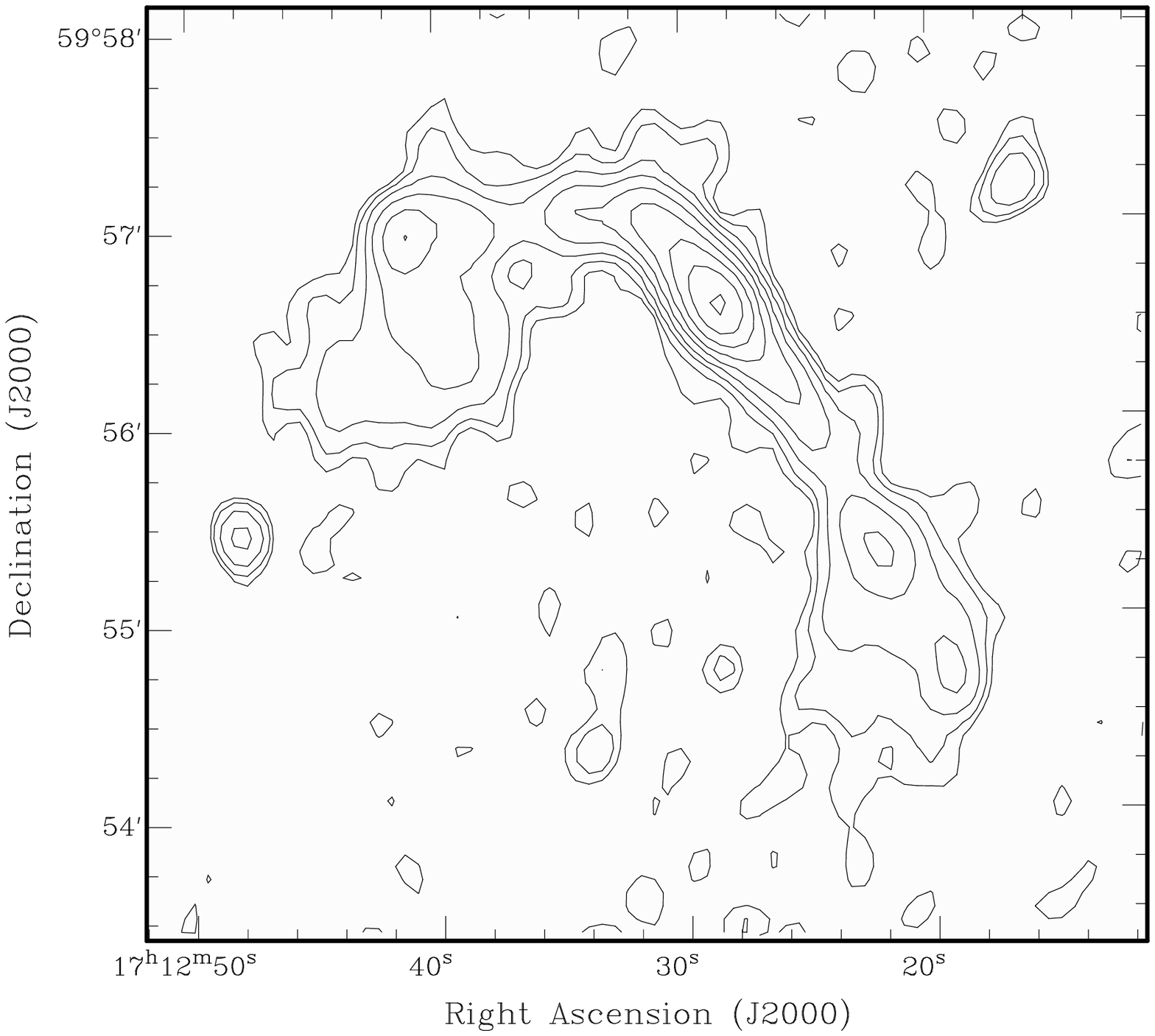,angle=0,width=5.5cm}
   \psfig{figure=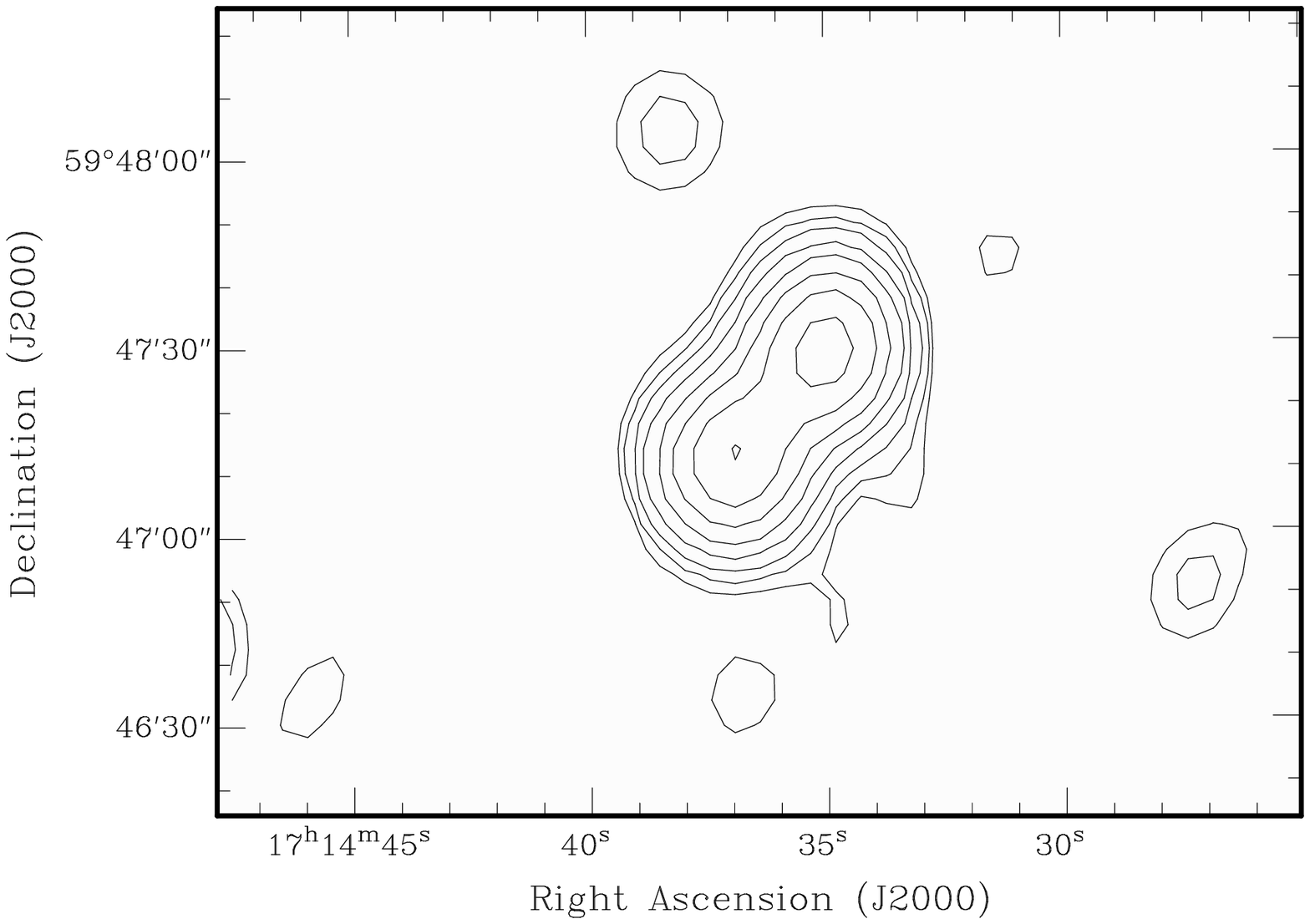,angle=0,width=5.5cm}
   \psfig{figure=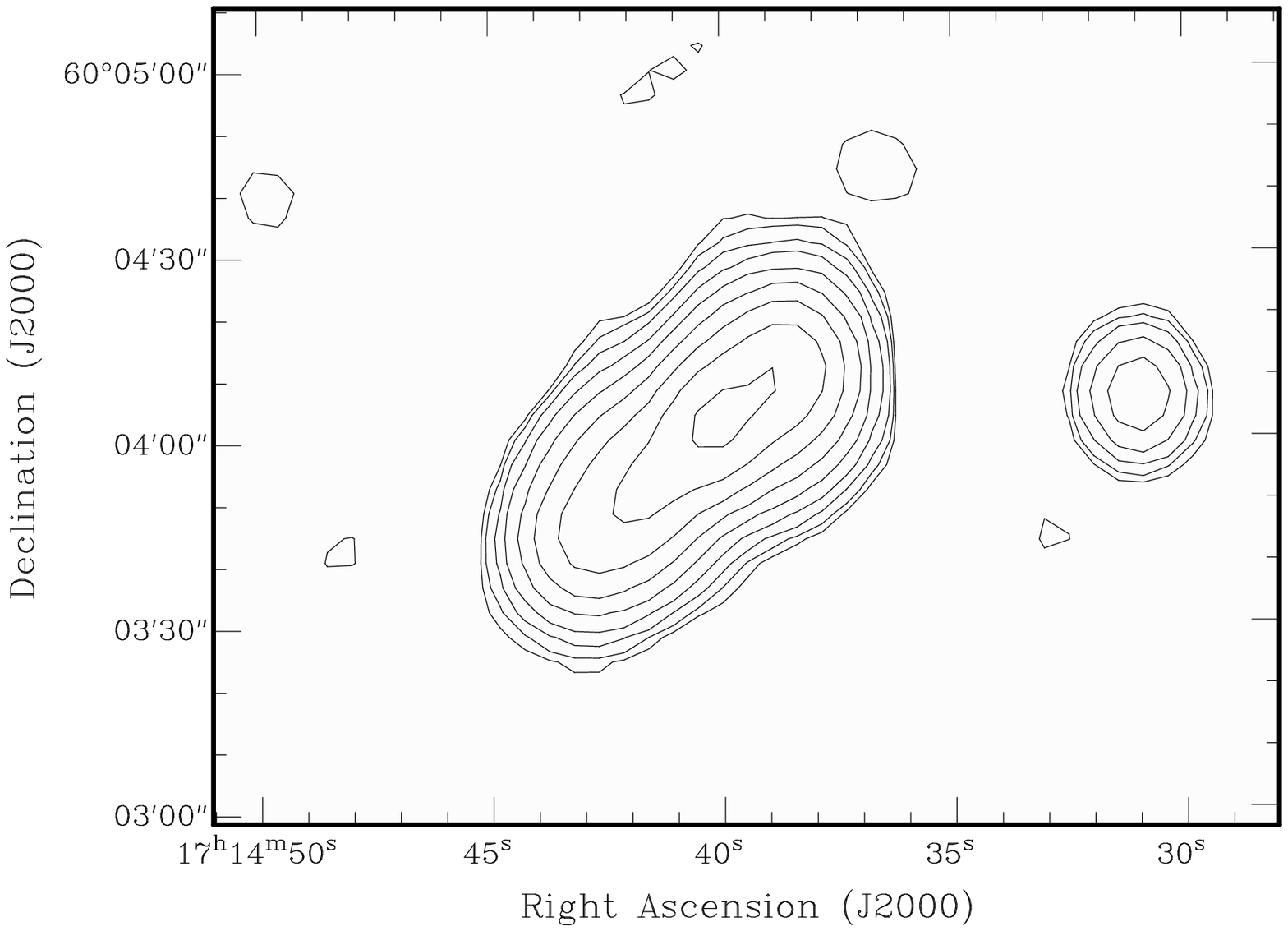,angle=0,width=5.5cm}}
\centerline{\psfig{figure=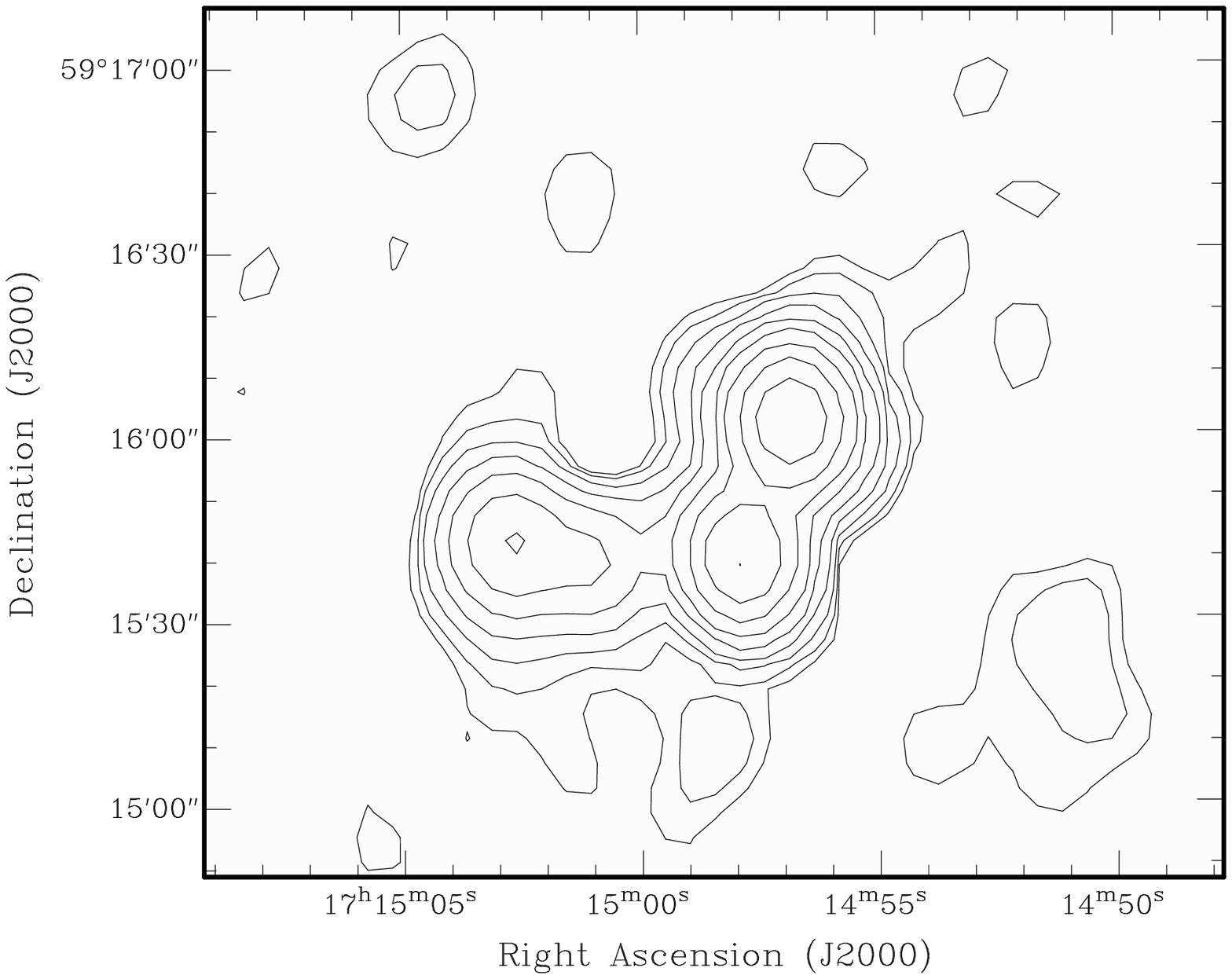,angle=0,width=5.5cm}
   \psfig{figure=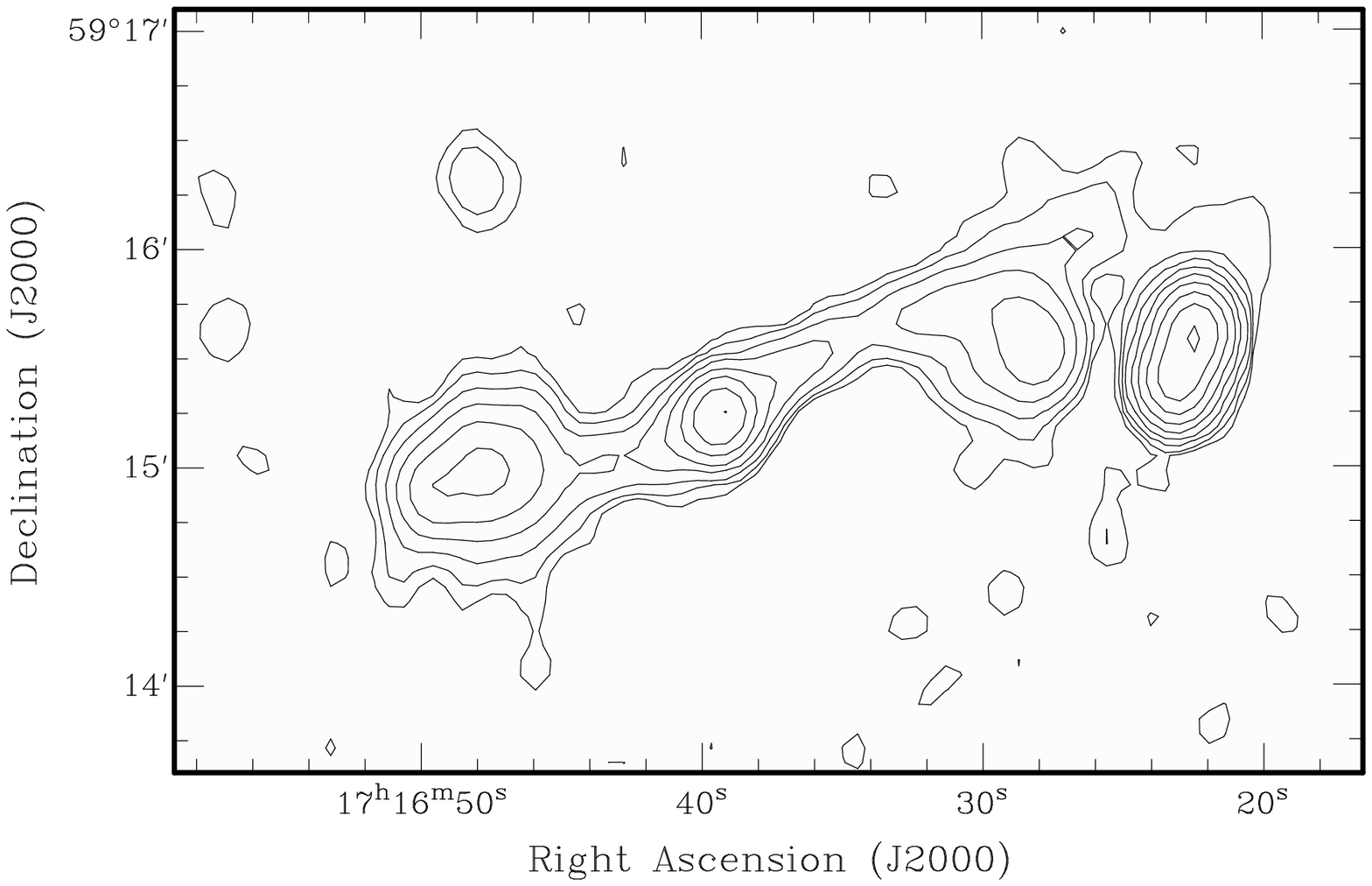,angle=0,width=5.5cm}
   \psfig{figure=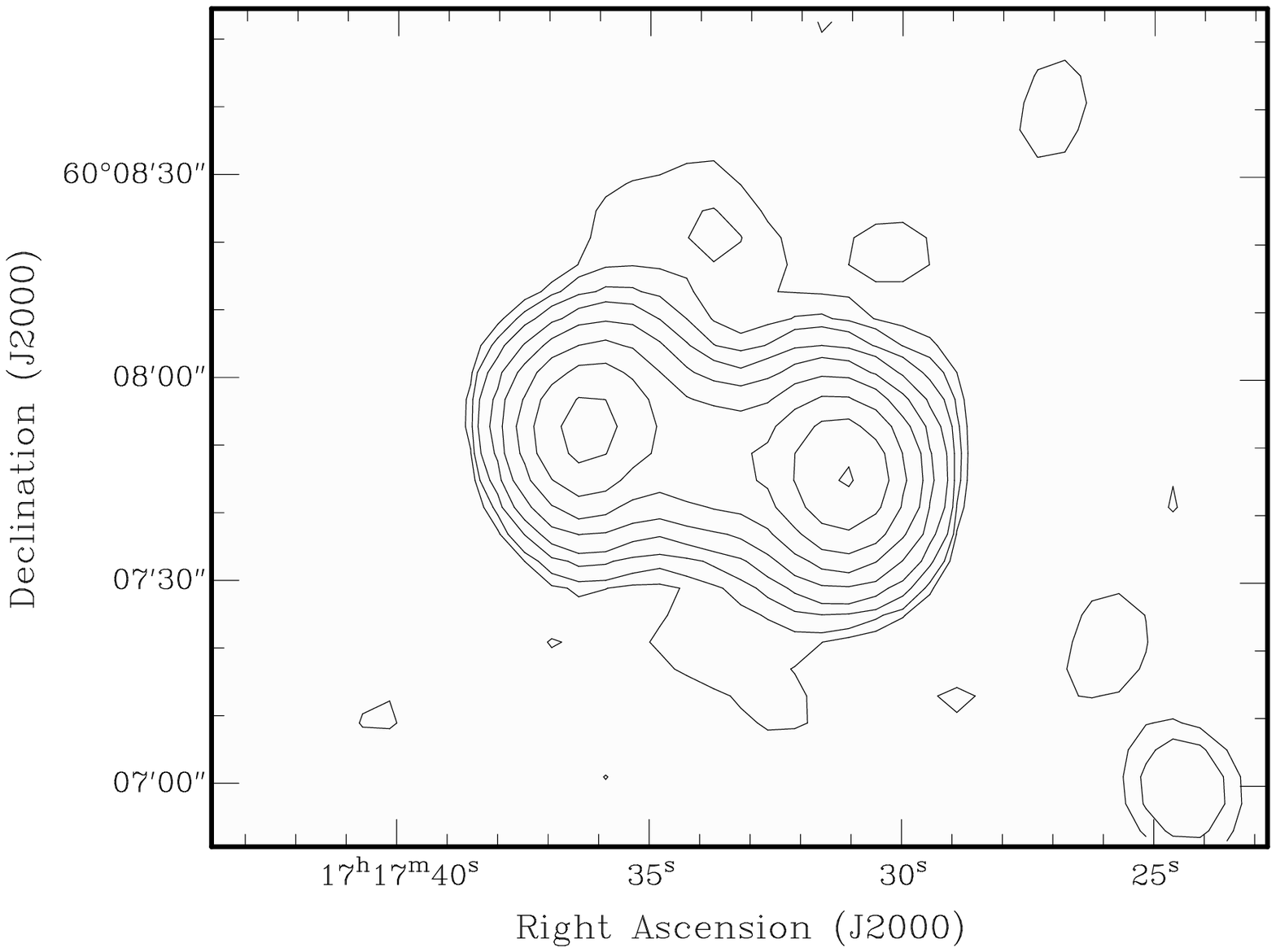,angle=0,width=5.5cm}}
\centerline{\psfig{figure=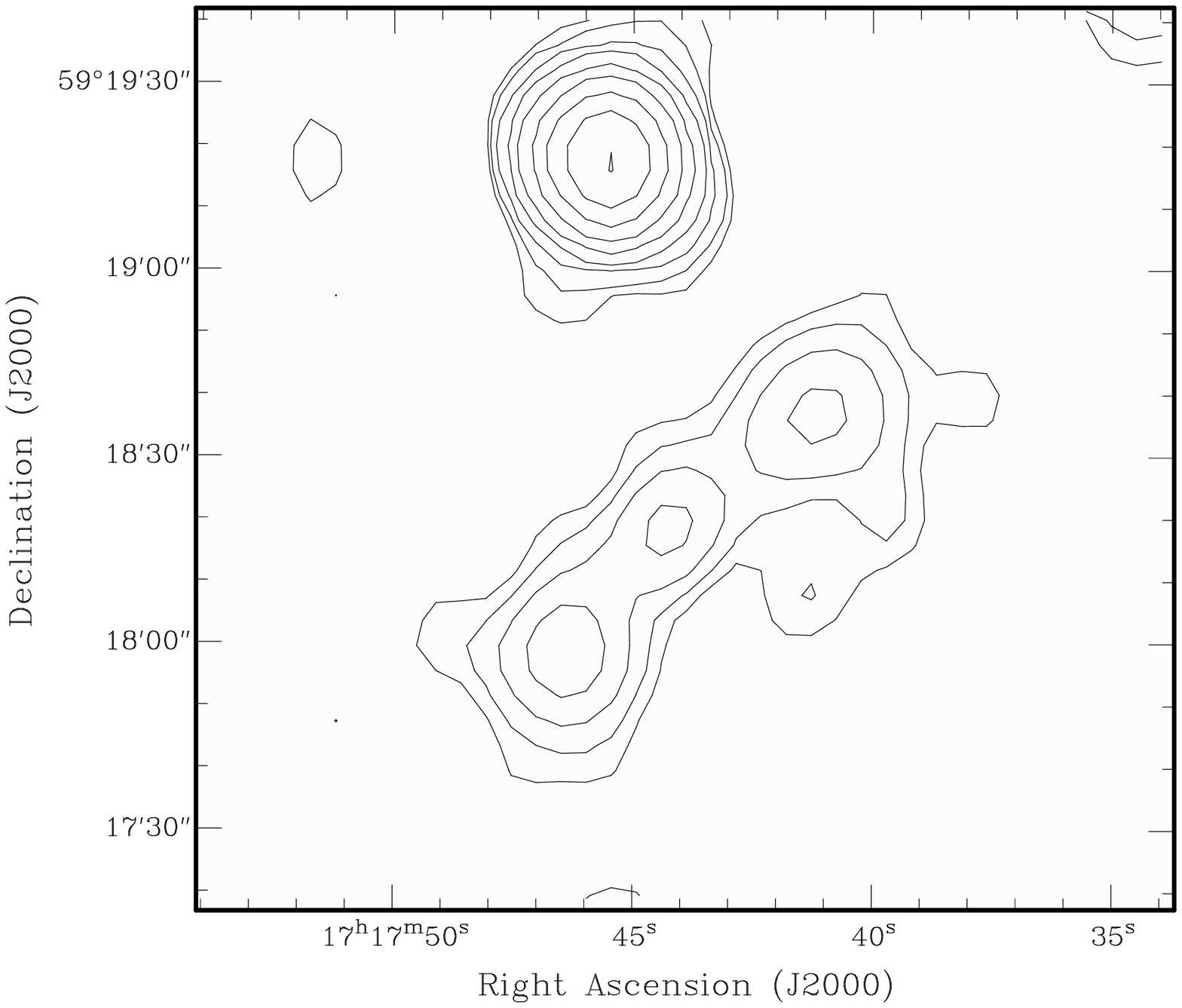,angle=0,width=5.5cm}
   \psfig{figure=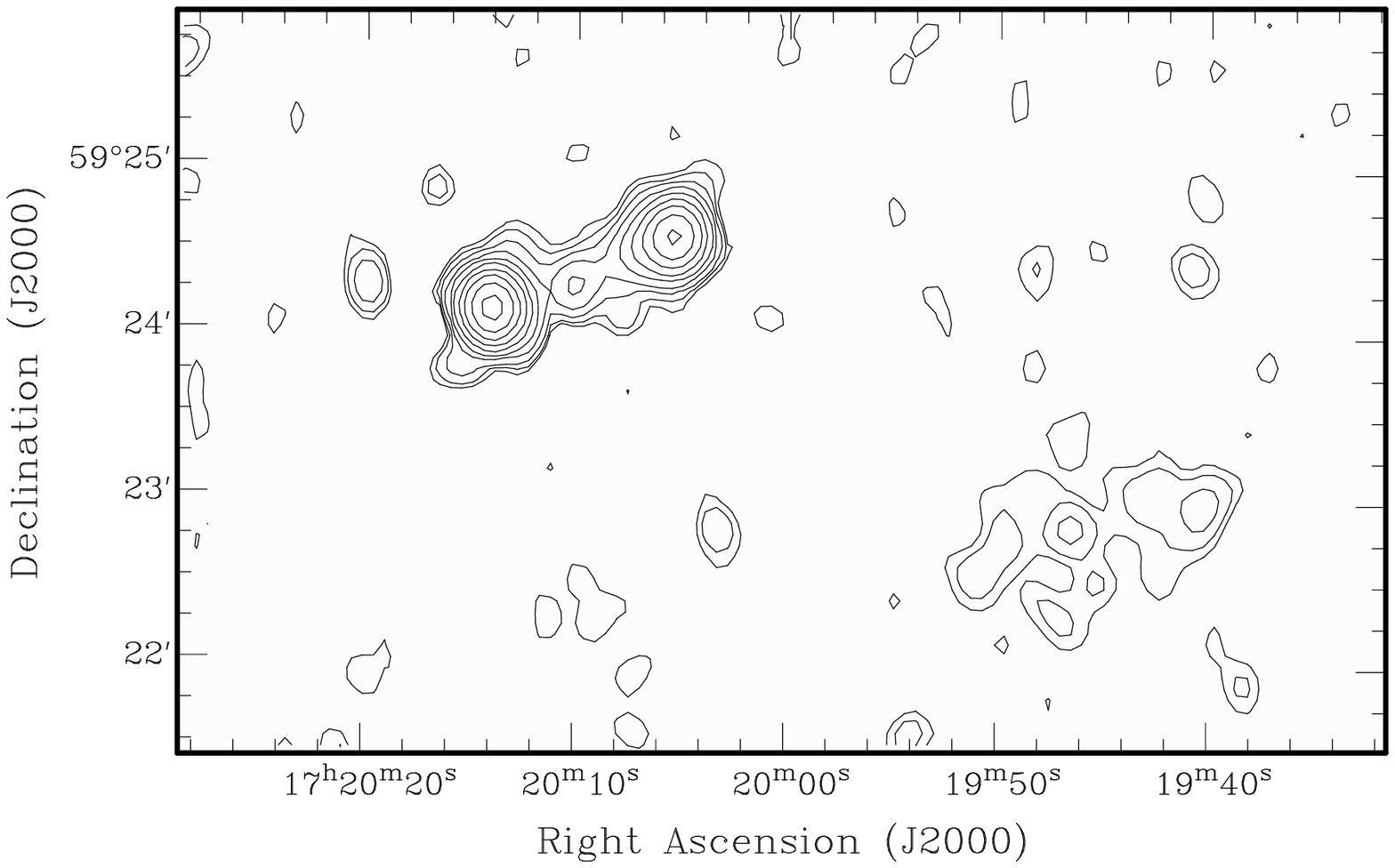,angle=0,width=5.5cm}
   \psfig{figure=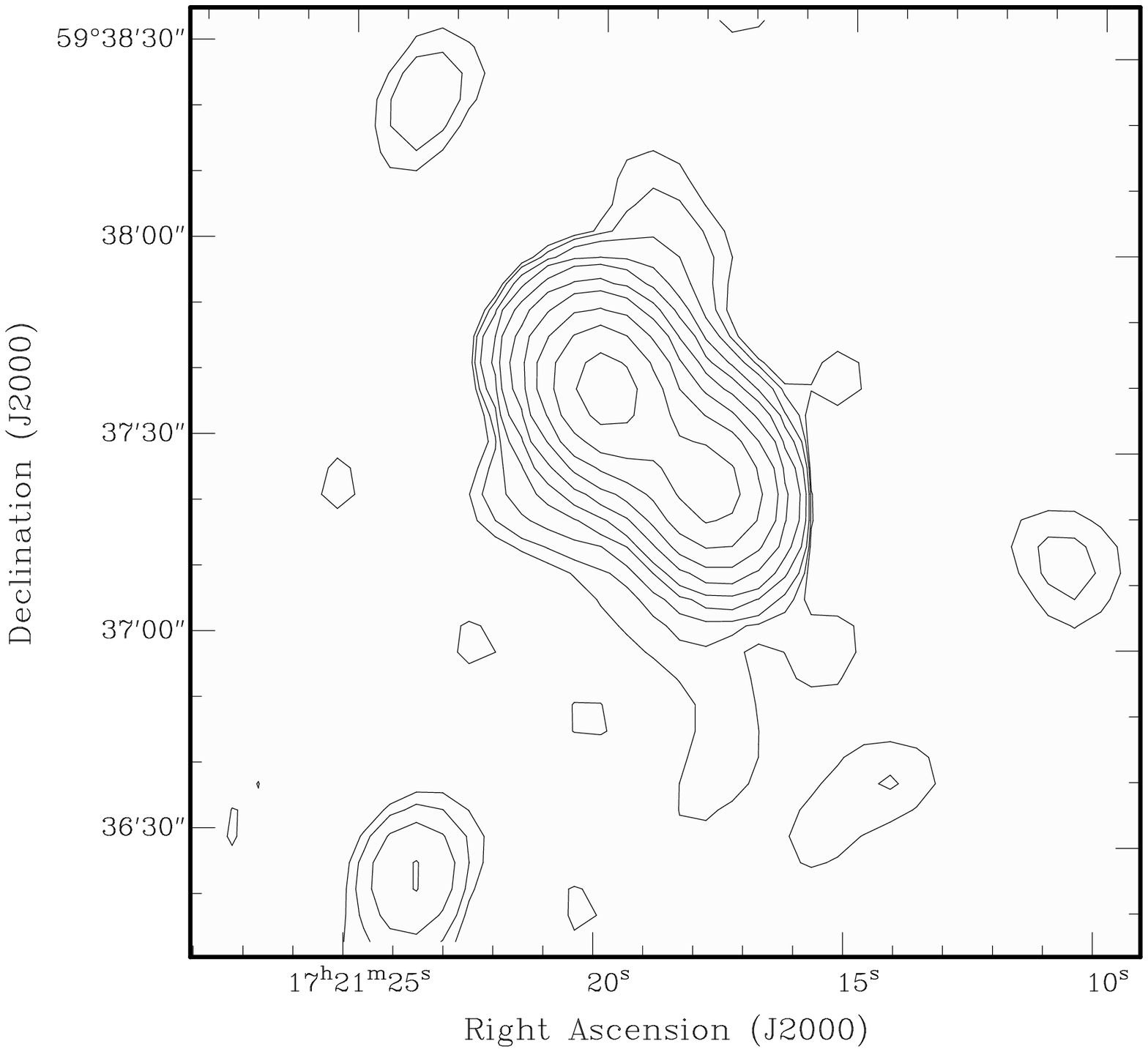,angle=0,width=5.5cm}}
\centerline{\psfig{figure=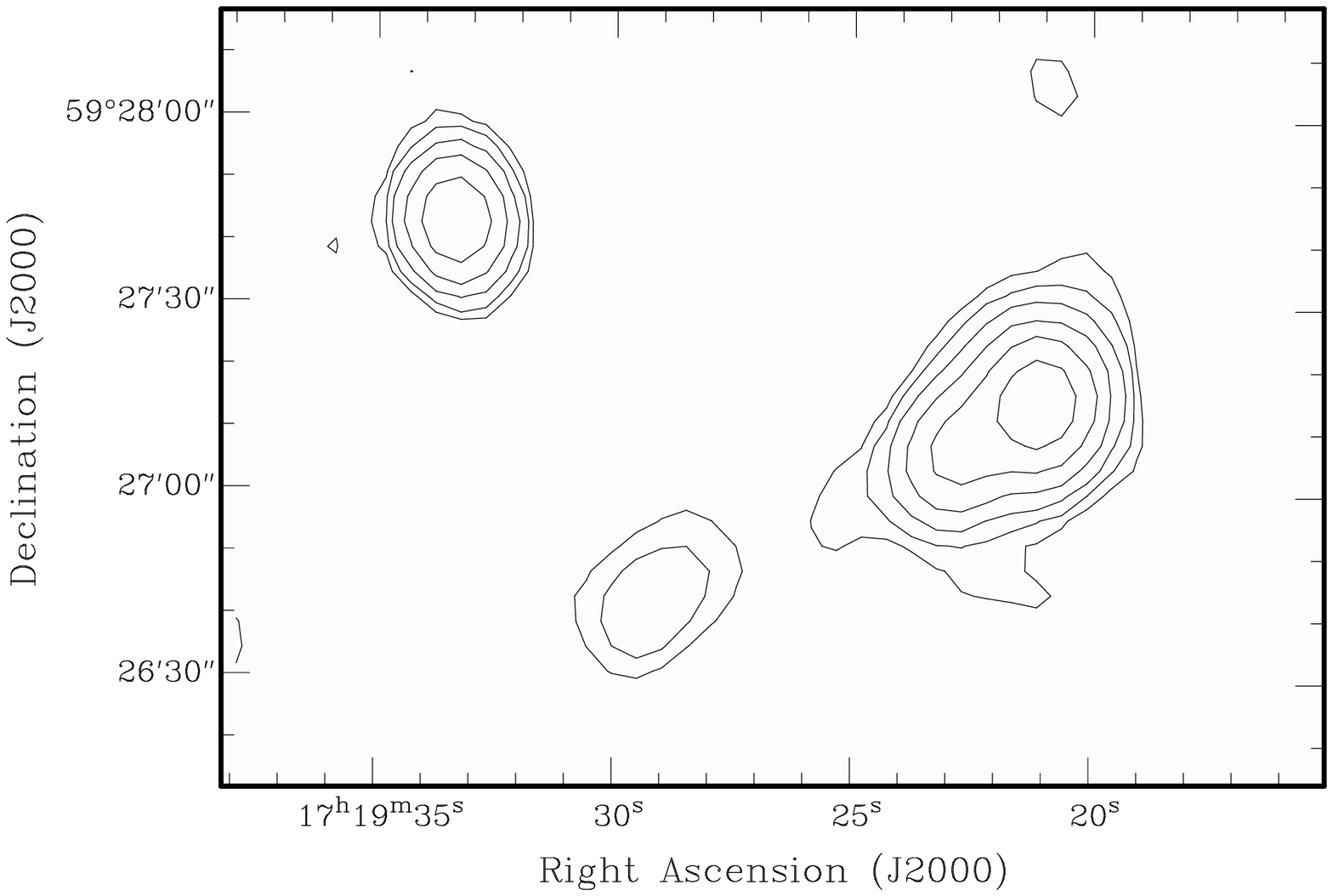,angle=0,width=5.5cm}
   \psfig{figure=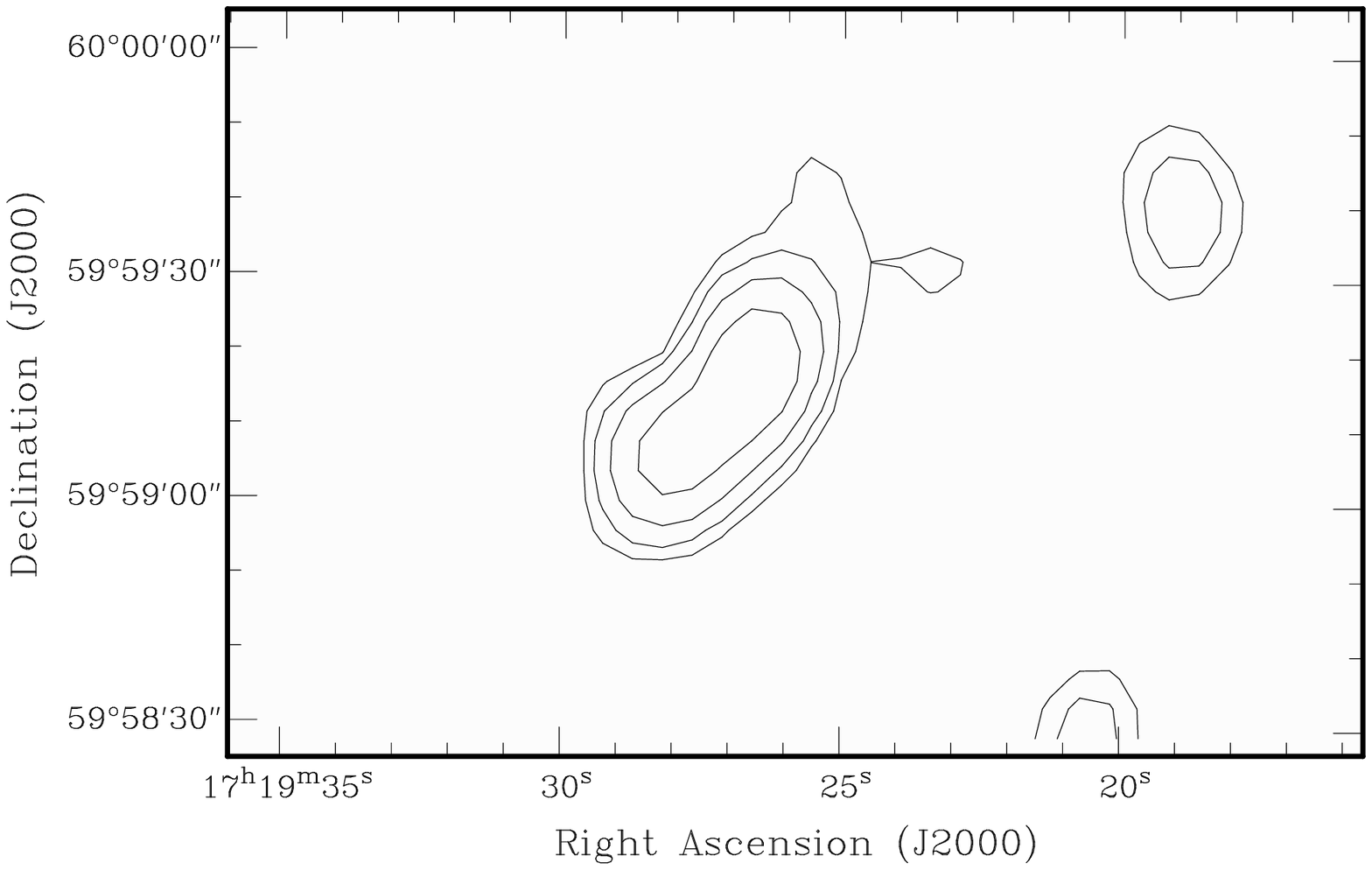,angle=0,width=5.5cm}
   \psfig{figure=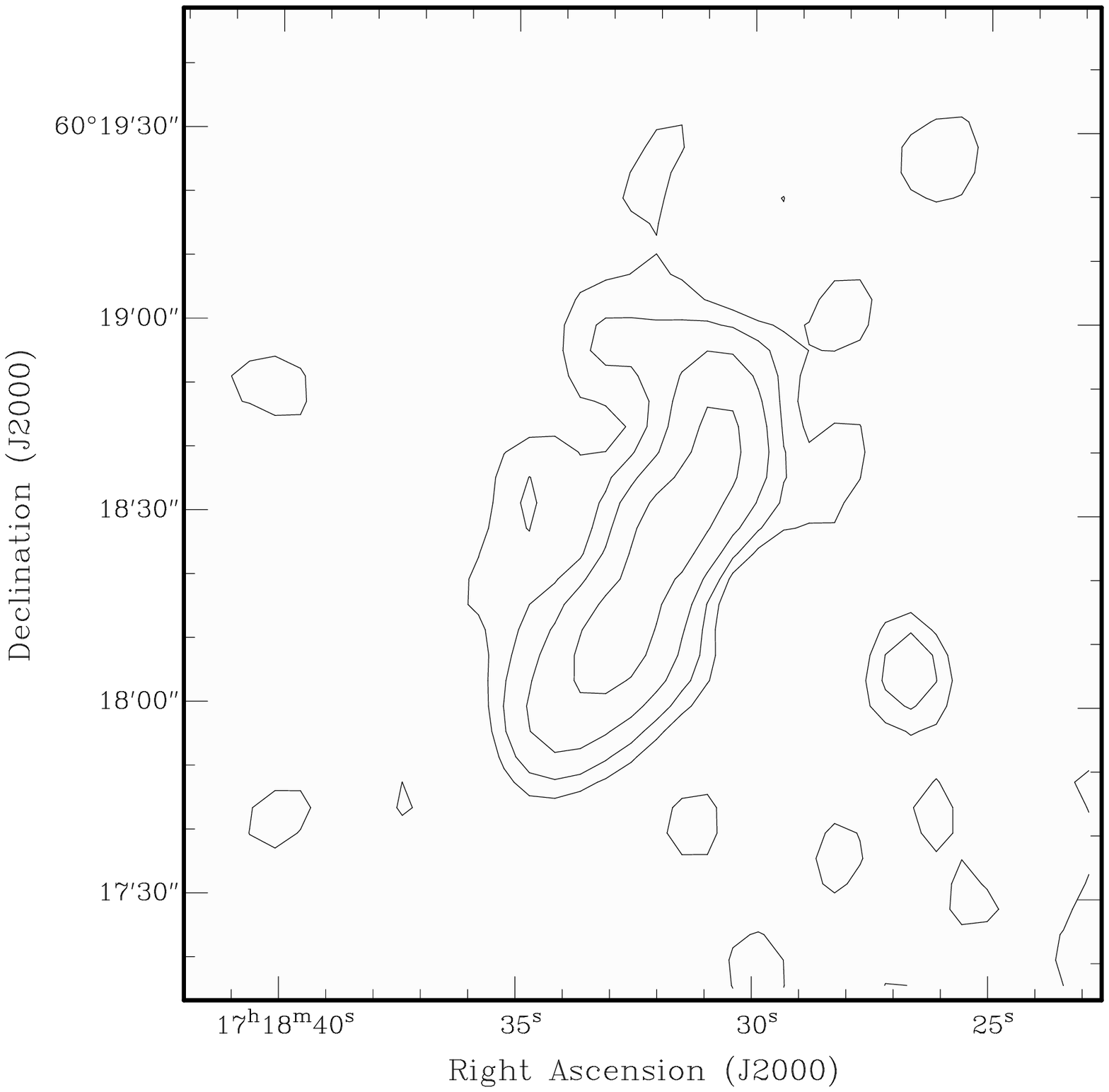,angle=0,width=5.5cm}} 
\caption{Examples of clearly resolved sources in the WSRT mosaic. 
   Contour levels are 30, 60, 120, ....  \microJy.  With the exception of the
   last three, all sources appear to be extended also in the VLA catalogue of
   Condon et al.  2003 (see also their Figs.  5 to 13).}
\end{figure*}

\section{Results}

The full mosaic and a more detailed view of the central region are shown in
Fig. 3. The final rms noise is $\sim 8.5$ \microJy\ in the central region
(increasing to $\sim 20$ \microJy\ at $\sim 30$ arcmin from the
centre). This is the deepest image produced with the WSRT at 1.4 GHz.

The automatic extraction of the sources was done using the algorithm SFIND 2.0
available inside MIRIAD. A full description of the algorithm as applied to
radio images, as well as a comparison with other algorithms, is given by
Hopkins et al.\ (2002). This routine incorporates a statistically robust
method for detecting sources, called the ``False Discovery Rate''. The details
of this method are given in Benjamini \& Hochberg (1995).

Here we only briefly summarise the SFIND 2.0 method. The algorithm
estimates the noise across the image. In our case the noise is mainly a
function of the radial distance from the centre. The values of the
local noise for each detected source are given in the last column of
Table 1. Each pixel is assigned a probability that it is drawn from the
background noise. A cutoff threshold on this probability is then
estimated, based on the percentage (chosen here to be 0.1\%) of false
detections of source pixels. The sources are then selected on this base
and their position, sizes, peak and total flux density are then
measured.  To produce the catalogue presented here, we have used
  the option of selecting sources for which the peak-pixel is above the
  estimated threshold.
  
The extraction of the sources was done in the inner 1 sq.  deg area (see
Fig.~2).  In this area 1048 sources were detected. The estimated background
noise reaches $\sim 8$ \microJy\ or better in the central region, comparable
to the expected theoretical noise ($\sim 6.5$ \microJy for uniformly weighted
data). The variation in the background noise level, shows the expected trend
to rise with increasing distance from the centre.  At 30 arcmins from the
centre of the field, the measured noise level increases to $\sim 20$
\microJy. An analysis of the ratio between the peak flux density of the source
and the background noise estimated around that source, shows that we are
reaching what would correspond to a $\sim 5 \sigma$ detection level in our
catalogue.  The number of detected sources compares well with that expected
from the known source counts (see e.g. Richards et al.  1998). A simple
estimate, considering the characteristics of the noise in our image
(increasing from the centre to the outer regions), indicates that we should
expect about 1150 sources at the $5\sigma$ level. The remaining discrepancy
can be due to the fact that some of the sources will be resolved by VLA
observations into multiple components.

For the detected sources, SFIND 2.0 provides as output a number of
parameters. They are listed in Table\ 1, where part of the catalogue is
presented to illustrate the format. The full version of the catalogue, as well
as the complete mosaic (in FITS format), can be down-loaded through the
WSRT-FLSv page at://www.astron.nl/wsrt/WSRTsurveys/WFLS.

We note in passing, that the more ``classical'' algorithm for source
extraction, IMSAD, was also tried on the FLSv field.  This algorithm (also
available in MIRIAD, see Prandoni et al. 2000 for details) selects all the
connected regions of pixels above a given flux density threshold (taken to be
5-$\sigma$ for the mosaic field of the FLSv). These regions are taken as the
sources present in the image above a certain flux limit. The algorithm then
performs a two-dimensional Gaussian fit of the flux distribution for each
source.  Although the catalogues produced with the two different methods
widely overlap, IMSAD proved to be less robust, compared to SFIND 2.0, against
false detections and also more prone to fail to detect group of sources when
they are separated by about a beam size.

Finally, there are a few sources that are clearly resolved with the $\sim
14^{\prime\prime}$ WSRT beam. Some examples of well resolved sources are
presented in Fig.\ 4.

\section{Confusion}
The position of the faintest sources in the catalogue may be affected by the
effects of source confusion. The usual rule of thumb is that confusion becomes
a problem when the source counts reach 1/30 per resolving beam. At this level
the positions, fluxes and sizes of the faintest sources deviate from their
true position since the measurement is also affected by the random
distribution of still fainter, unresolved sources that are not detected
individually. This sea of unresolved background sources can therefore modify
the measurements for the faintest sources in our catalogue. Hogg (2001)
simulated images of crowded fields, assuming various slopes for the
differential source counts. Following Hogg's definition, the source counts in
our survey reach about 1/44 per resolving beam. Taking a slope of -1.5
(somewhat steeper than the measured slope e.g. Richards 2000), we estimate
from Hogg's Figure 4, that due to the effects of source confusion, the median
error in the measured positions of our faintest sources will typically be
$\sim 0.2$ of the FWHM of WSRT resolving beam. Occasionally (about 10\% of the
time) sources will have larger errors, about 0.4 of the FWHM. Thus for the
faintest sources in our catalogue, positions of some sources could, in
principle, be in error by as much as 6 arcseconds. In a similar way, confusion
can also affect the measured source sizes. Flux densities errors can also
arise, the median error is $\sim 5$\% but occasionally these can be as much as
30\%.
  
It should be noted that the errors induced by source confusion are confined to
the deepest, central areas of the WSRT image.  At 15 arcminutes from the
central field, the noise rises to 11 \microJy, the source counts fall to $\sim
1/63$ per resolving beam, and the effects of confusion are very much
reduced. At about 22 arcminutes from the central field the noise rises to 15
\microJy, the source counts fall to $\sim 1/100$ and the effects of confusion
become negligible.
  
The WSRT FLS FITS file is in the public domain. Astronomers using this image
to cross-identify sources at the 3 sigma level, should be aware that the
effects of confusion on estimated positions, fluxes etc. will be much larger
(the source counts rise to $\sim 1/22$ per resolving beam in the central part
of the field), making detailed comparisons unreliable.

\section{Positional Accuracy}

As we have discussed in the previous section, the positional errors
associated with the faintest sources in our catalogue are likely dominated by
the effects of source confusion. However, for the brighter sources it is still
important to compare the positions of our extracted sources with those from
the VLA catalogue of Condon et al.  (2003). Given the difference (more than a
factor 2) in resolution, this comparison is not completely straightforward. In
particular, the number of resolved sources is much higher for the VLA
observations, and therefore the offset between the two may reflect the
different morphologies of the sources at the two resolutions.

In Fig.\ 5 we show the distribution of the offset in RA and DEC as obtained
from the large (1-degree) region.  The size of the symbols is proportional to
the peak flux.  Similar results are found when comparing the sources in the
inner (25-arcmin) region. The median offset is about 1 arcsec. As expected,
weaker sources tend to have a larger offset. Following the relation discussed
in Prandoni et al. (2000) and Condon (1998), the formal positional accuracy
expected in the case of our observations is $\sim 1.5^{\prime\prime}$ for a
point source at the limit of the survey (5 $\sigma$).


However, as shown from the cross plotted in Fig.\ 5, a systematic offset of
about 0.5 arcsec has been found in declination.  In RA the offset is less than
0.1$^{\prime\prime}$.  This offset in declination corresponds to less than a
1/20 of the synthesised beam width.  Nevertheless, given that this offset
seems to be independent of the flux density  of the sources, it probably
reflects the typical astrometric accuracy the WSRT can achieve, using standard
calibration techniques.

No correction for the offset has been applied to the coordinates
presented in the catalogue.

\begin{table*}
\centering
\caption{\HI\ detections. The \HI\ masses are estimated assuming $H_\circ = 70$ \kms Mpc$^{-1}$. }
\begin{tabular}{lccccc}
\hline\hline\\
Name & RA & DEC  & Vel  & \HI\ Mass & S$_{1.4{\rm GHz}}$ \\
     & J2000  & J2000 & \kms\   &  M$_{\odot}$  &  $\mu$Jy          \\
\hline \\
J171441+595052  & 17:14:41 & 59:50:52 & 5474 & $2.1 \times 10^9$   & 87  \\
J171603+592333  & 17:16:03 & 59:23:33 & 7881 &  $2.6 \times 10^9$  &  -- \\
J171612+594026  & 17:16:12 & 59:40:26 & 16921 & $6.1 \times 10^9$   & 43 \\
J171729+594757  & 17:17:29 & 59:47:57 & 23551 & $1.3 \times 10^9$   &  --\\
\hline 
\end{tabular}
\end{table*}

\section{A Serendipitous Search for \HI\ Emission}

As described in section 2, the data presented here included observing
bands that were set to frequencies below 1421~MHz, the rest-frequency
of neutral hydrogen emission. Indeed, the standard continuum WSRT
set-up is well suited for a serendipitous search for \HI\ emission
(and also \HI\ absorption, if strong sources are present in the field).
The frequency range of our observations, together with the large number
of channels employed for each 20~MHz band, thus permitted us to search
for serendipitous \HI\ emission from galaxies in the FLSv field.

We selected a subset of the observations, specifically the $2 \times 12$ hours
centred on the FLSv, to look for possible serendipitous \HI\ emission. After
subtracting the continuum CLEAN components from the (self-calibrated)
visibility data sets, the data were Hanning smoothed to suppress Gibbs effects
that are inherent to digital correlators.

A region covering about 40$^{\prime\prime}$ on a side was imaged using a
robust weighting equal to 1.  The synthesised beam is $24^{\prime\prime}
\times 20^{\prime\prime}$ (in PA=1.4$^\circ$).  Of this region, a
spectral-line data cube of 410 channels with a velocity resolution of 60 \kms\
width was generated.  It should be noted that a resolution of 60 \kms\ is good
enough to distinguish between single or multiple galaxies, and even measure
the galaxies rotation velocity for massive \HI\ systems.  A second data set,
with a velocity resolution further smoothed to 100 \kms\ was generated for
comparison purposes.  The entire data cube for both data sets covered the
velocity range from 0 to 24600
\kms.  The typical ($1\sigma$) noise level obtained in each 60 \kms\ channel
was about 0.12 mJy/beam.

\begin{figure}
\centerline{\psfig{figure=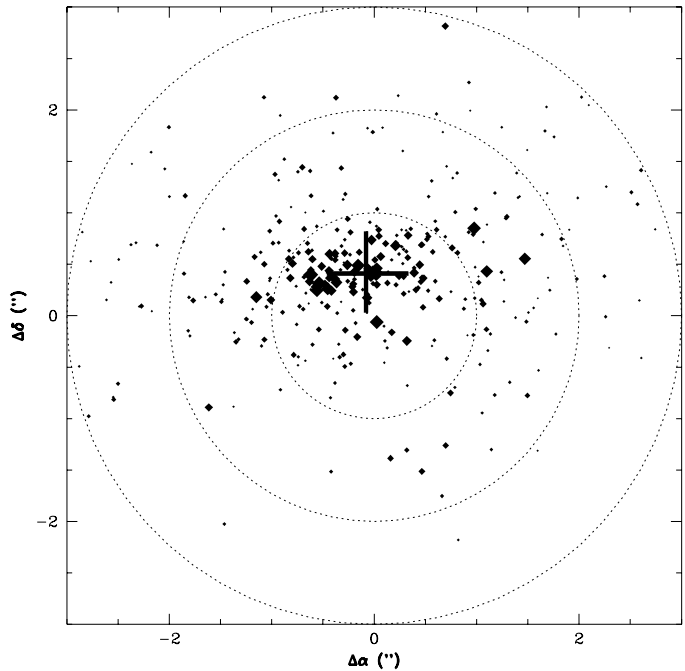,angle=-90,width=9.0cm}}
\caption{Distribution of the position offsets (in right ascension and
declination)  between the sources in common to
the WSRT and the VLA in the 1-degree region of the WSRT mosaic. The size of
the symbols is proportional to the flux of the source.}
\end{figure}

The final data cube was searched for \HI\ emission using both a visual
inspection of the data and an automatic detection routine (kindly provided by
T. Oosterloo).  The candidate detections were then further inspected for
optical counterparts using deep optical images obtained with the NOAO
telescope (5$\sigma$ depth limiting magnitude of R = 25.5 (Vega), Fadda et
al. 2004) and publicly available at
http://ssc.spitzer.caltech.edu/fls/extragal/noaor.html.  Four
\HI\ detections were found in both the 60 and 100 \kms\ data cubes.  Fig.  6
shows these results, with the total \HI\ intensity of the confirmed detections
superimposed upon the optical image.  The parameters of the
\HI\ detections are summarised in Table 2.  All four \HI\ detections
are identified with nearby, massive galaxies.  A similar number of
serendipitous \HI\ detections were also found (using the same technique) in a
deep survey (of similar extent) performed within the NOAO-N ``Bootes'' field
(Morganti \& Garrett 2002).  The results of a search
for serendipitous \HI\ detections from the full mosaic data set, will
be presented in a forthcoming paper. No \HI\ absorption was detected
towards any of the sources in the 40$^{\prime\prime}$ FLSv field.

\begin{figure*}
\centerline{\psfig{figure=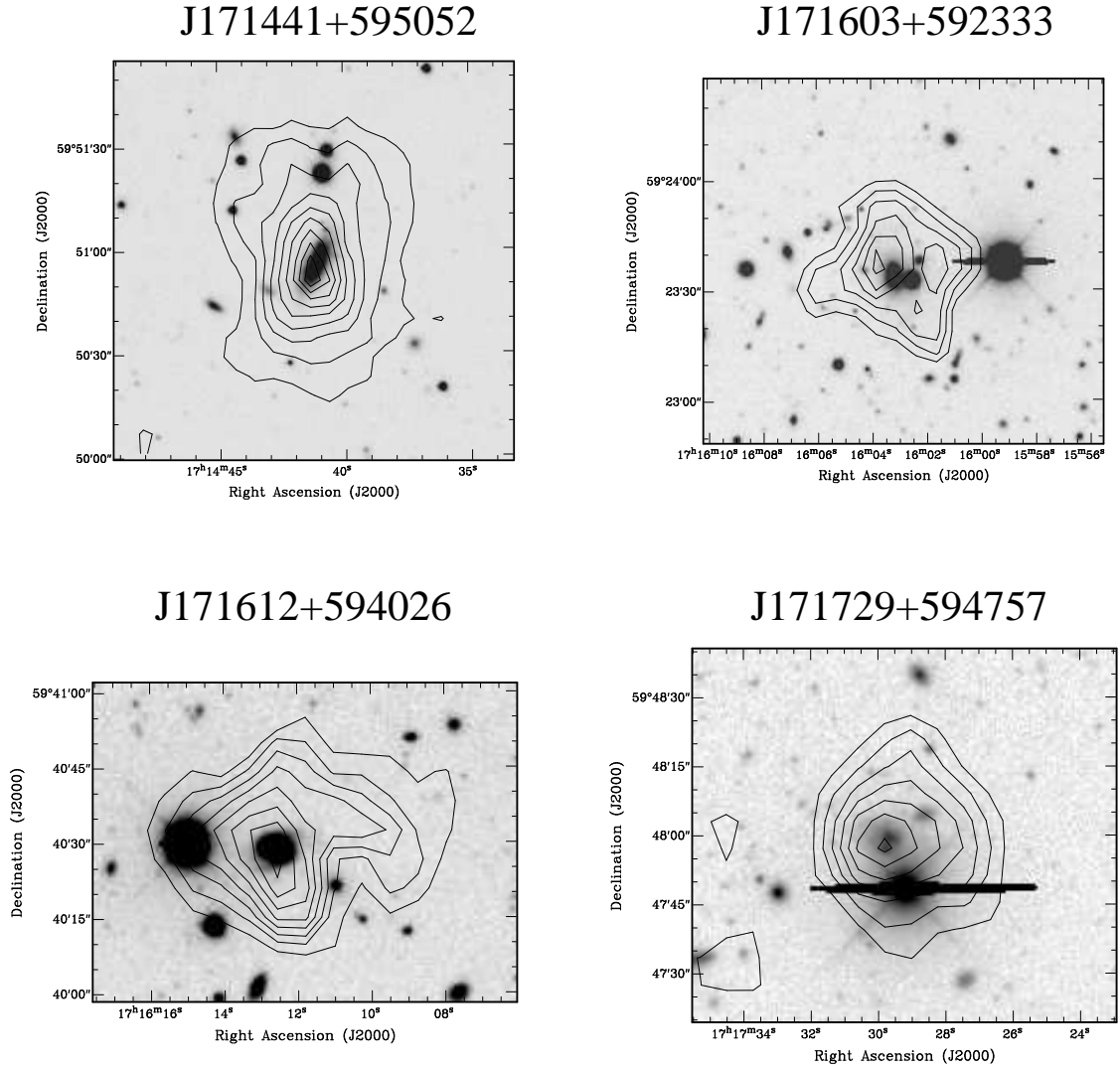,angle=0,width=16cm}}

\caption{\HI\ total intensity contours 
superimposed on optical images of the four \HI\ detections in the central
region of the FLSv field. The contour levels are: 
for J171441+595052
$2.2 \times 10^{20}$ to  $2.9 \times 10^{21}$ cm$^{-2}$ 
in steps of $3.2 \times 10^{20}$ cm$^{-2}$; 
for J171603+592333
$4.0 \times 10^{20}$ to  $1.6 \times 10^{21}$ cm$^{-2}$ 
in steps of $2.2 \times 10^{20}$ cm$^{-2}$;
for J171612+594026
$1.8 \times 10^{20}$ to  $7.0 \times 10^{20}$ cm$^{-2}$ 
in steps of $7.3 \times 10^{19}$ cm$^{-2}$;
for J171729+594757
$1.8 \times 10^{20}$ to  $1.4 \times 10^{21}$ cm$^{-2}$ 
in steps of $1.7 \times 10^{20}$ cm$^{-2}$;}
\end{figure*}

Finally, an estimate using the \HI\ mass function (Zwaan 2000) suggests that
every field observed with a similar depth should contain about 5 objects
with detectable \HI\ emission.  The majority of the expected objects will be
$M_\star$ galaxies, i.e. galaxies with an amount of neutral hydrogen between
$10^9$ to $10^{10}$ \msun.  The bias toward the detection of massive systems,
is partly due to the relatively coarse velocity resolution that is obtained
using the ``default" continuum set-up.

\section{Follow up observations}

Compared to other deep field radio observations (e.g. the much smaller
regions covered by the Hubble Deep Field surveys), the VLA and WSRT-FLSv
catalogues cover significant areas of sky with excellent microJy
sensitivity. In particular, the deep radio observations presented here,
will ensure significant overlap between the radio and FIR Spitzer FLSv
observations out to at least $z\sim 1$. Hyper-luminous star forming
galaxies will be detected at even higher redshifts. 

Spectroscopic and photometric redshift campaigns in the region of the FLSv are
already underway. Successful SCUBA observations have been carried out using a
radio selection combined with optical red colours (Frayer et al. 2004). The
complementarity between the Spitzer and WSRT observations will enable the
first detailed study of the FIR-Radio correlation to be made, out to at least
$z\sim 1.5$. While there is already good evidence that the correlation
persists out to at least $z\sim 1$ (Garrett 2002), the Spitzer and WSRT
observations may uncover more subtle deviations or possible large scale trends
in the relation, as a function of redshift or galaxy (starburst) type. In
addition, sources that deviate from the relation will be easily identified as
rare and interesting AGN, in a radio and FIR field that is otherwise dominated
by star forming systems.

These wide area VLA and WSRT radio surveys will also be important in
acting as a finding chart for very deep, high resolution, wide-field
VLBI observations of the FLSv region. By using the technique of
full-beam calibration (Garrett et al. 2003), it should be possible to
image several dozen faint (radio-loud) AGN in the field with full
microJy sensitivity and milliarcsecond resolution. For the first time,
it will be possible to conduct a large-scale census of the
milliarcsecond structures of these sub-mJy and microJy radio sources.
Astrometric positions of the sources detected will be measured with
reference to the IERS reference frame with a precision of a
milliarcsecond or better. 
The first results are already presented in Wrobel et al. 2004. The VLBI observations will also provide an
independent and unambiguous check on the AGN content of the field.

Finally, we note in passing that even deeper serendipitous \HI\ observations
are possible with the WSRT's standard L-band continuum set-up.  In particular,
the 21-cm band can reach $\sim 1160$ MHz (albeit with a less favourable RFI
environment).  By selecting frequencies at the edge of the band, one can, in
principle, explore serendipitous \HI\ emission up to redshift of $z = 0.22$.

\begin{acknowledgements}
We thank T. Oosterloo for his help with the \HI\ data and in particular for
providing the routine for the automatic detection of \HI\ candidates.
The authors also thank the referee for many useful comments that have helped
to improve the content of the paper.
The WSRT is operated by the ASTRON (Netherlands Foundation for
Research in Astronomy) with support from the Netherlands Foundation
for Scientific Research (NWO).
\end{acknowledgements}

\end{document}